\documentclass[twocolumn,amsmath,amssymb,superscriptaddress]{revtex4}
\usepackage{graphicx, color}
\graphicspath{{figures/pdf/}}
\DeclareGraphicsExtensions{.pdf}
\usepackage{dcolumn}
\usepackage{bm}
\usepackage{amsthm,amsmath,amssymb}
\usepackage{verbatim}
\usepackage{hyperref}
\usepackage{fontenc}
\usepackage{cfkc}

\def\Re{\mathrm{Re}}
\def\Im{\mathrm{Im}}
\def\ket#1{| #1 \rangle}
\def\bra#1{\langle #1 |}

\def\RR{\mathbb{R}}
\def\diag{\operatorname{diag}}

\def\D{\mathcal{D}}
\def\BB{\mathfrak{B}}
\def\LL{\mathfrak{L}}
\def\H{\mathcal{H}}
\def\ONE{\mathbb{I}}

\begin{document}

\title{Fundamental Speed Limits on Quantum Coherence and Correlation Decay}

\author{Daniel K. L. Oi} 
\affiliation{SUPA Department of Physics,
  University of Strathclyde, Glasgow G4 0NG, 
  United Kingdom}\email{daniel.oi@strath.ac.uk}

\author{Sophie G. Schirmer} 
\affiliation{Department of Applied Maths and Theoretical Physics, 
  University of Cambridge, Wilberforce Road, Cambridge CB3 0WA, 
  United Kingdom}

\affiliation{College of Science (Physics), Swansea University,
  Singleton Park, Swansea, SA2 8PP, United Kingdom}

\email{sgs29@cam.ac.uk, sgs29@swan.ac.uk}

\date{\today}

\begin{abstract}
  The study and control of coherence in quantum systems is one of the
  most exciting recent developments in physics. Quantum coherence
  plays a crucial role in emerging quantum technologies as well as
  fundamental experiments. A major obstacle to the utilization of
  quantum effects is decoherence, primarily in the form of dephasing
  that destroys quantum coherence, and leads to effective
  classical behaviour. We show that there are universal relationships
  governing dephasing, which constrain the relative rates at which
  quantum correlations can disappear. These effectively lead to speed
  limits which become especially important in multi-partite systems.
\end{abstract}

\maketitle

One of the principle distinguishing features between classical systems
and quantum systems is the existence of quantum coherences leading to
correlations that cannot be accounted for classically. For example,
the phenomenon of entanglement~\cite{Schroedinger1936} and the
violation of non-local realism~\cite{Bell1964} are such consequences.
The manipulation and preservation of such coherences is vital for
tasks such as the construction of quantum information processing
devices~\cite{Deutsch1985}, quantum communication~\cite{Teleport1993},
cryptography~\cite{Ekert1991}, and metrology~\cite{Caves1981}. It has
also been suggested that quantum coherence plays a role in certain
biological processes~\cite{QBio2007}. Unfortunately, the inevitable
interaction of quantum systems with the environment leads to
decoherence, the dominant form of which is dephasing, or the
disappearance of the off-diagonal elements of the density operator of
the system~\cite{Zurek1981}.  The rates at which these coherences decay are
crucial as they determine how quickly the system approaches
classicality, a state where quantum correlations are
lost. Considerable effort has been expended on understanding the
fundamentals of decoherence and quantum correlations, and how to
protect the latter and prevent rapid decay using encodings in
protected subspaces or subsystems, for example~\cite{Palma1996}.

Previously, some surprising relationships between the dephasing rates in
multi-level quantum systems have been
uncovered~\cite{SS2004,Berman2005}.  These stem from the need to
preserve positivity of the density operator, or more generally the
complete positivity of the evolution~\cite{Kraus1971}, and lead to
constraints on the relative rates of dephasing.  The surface of this
phenomenon and has only been scratched, and in this work we elucidate
the universal nature of these constraints and present a general
framework. An important consequence of these constraints are effective
speed limits on the decay of correlations and entanglement in
multi-partite systems. These speed limits are independent of the details
of the Hamiltonian evolution or decoherence mechanisms and therefore
apply to a large class of quantum systems from nuclear spins, to atoms,
ions and quantum dots to biomolecules.

We consider an $N$-level quantum system subject to Markovian dephasing
whose evolution can be described by a Lindblad master
equation~\cite{Lindblad1976,GKS1976}.  A \emph{pure dephasing process}
leaves the populations of the basis states invariant, and leads to decay
of the magnitude of the off-diagonal elements (coherences), as well as
frequency shifts.  Previous work for three-level systems found that the
decay rates and frequency shifts were
constrained~\cite{SS2004,Berman2005}. Additionally, \cite{SS2004} also
gave some partial results for four-level systems and showed such
constraints must exist for higher-level systems but the equations were
intractable in general.

Here, we present a canonical form for the pure dephasing Lindblad
operators which allows us to derive a general form for the constraints
for $N$-level systems. These form a hierachy of inequalities, defining a
convex cone of allowed dephasing rates.  The general form also allows us
to invert physically observed dephasing rates to define a unique set of
canonical dephasing operators, which reflect correlations between noise
processes such as fluctuations in the energy levels, and may serve as a
useful diagnostic tool. In multi-partite systems these constraints
induce speed limits on the decay of non-local quantum correlations and
entanglement in terms of the local dephasing rates.

\section*{Results}

\subsection*{Canonical Dephasing Operators}

The key idea is that pure dephasing of an $N$-level
system may be modelled by a diagonal Hamiltonian $H=\diag(\lambda_n)$
and a canoncial set of $N-1$ or fewer diagonal Lindblad operators
$\{V_k\}_{k=1}^{N-1}$ of the special form
\begin{equation}
\label{eq:canonical}
  V_k = \diag \left(\underbrace{0 \ldots 0}_{k}, a_{k+1}^{(k)},
         \ldots a_{N}^{(k)} \right),
\end{equation}
where the non-zero diagonal elements $a_n^{(k)}$ can be complex except
for the first non-zero element, $a_{k+1}^{(k)}$, which is set to be
real and non-negative. The density operator elements evolve as
\begin{equation}
  \label{eq:rhomn}
  \rho_{mn}(t)= e^{-t(i\omega_{mn}+\Gamma_{mn})} \rho_{mn}(0),
\end{equation}
with effective frequencies given by $\omega_{mn}=\lambda_m-\lambda_n+
\Delta\omega_{mn}$, and dephasing-induced frequency shifts and
decoherence rates
\begin{subequations}
 \label{eqs:omgGamma}
\begin{align}
  \Delta\omega_{mn}
              &= -\sum_k \Im(a_{mk}a_{nk}^*), \\
  \Gamma_{mn} &= \tfrac{1}{2}\sum_k (|a_{mk}|^2+|a_{nk}|^2)
                 -\Re(a_{mk}a_{nk}^*).
\end{align}
\end{subequations}

The populations are constant as $\omega_{nn}=\Gamma_{nn}=0$.  The
off-diagonal elements decay with the damping rate $\Gamma_{mn}$
\begin{equation}
  |\rho_{mn}(t)| = e^{-t \Gamma_{mn}} |\rho_{mn}(0)|.
\end{equation}
If the $a_{nk}$ are real then the expressions simplify, $\Gamma_{mn}=
\tfrac{1}{2}\sum_k(a_{mk}-a_{nk})^2$, and there are no frequency shifts,
$\Delta\omega_{mn}=0$.

As shown in the Methods, any set of pure dephasing Lindblad operators
can be transformed to this form leaving the total superoperator
unchanged.  This reduces an arbitrary number of parameters, specified
by the non-zero elements of an arbitrary set of dephasing operators,
to $N(N-1)/2$ parameters in the canonical form.  The number of free
parameters matches exactly the number of dephasing rates $\Gamma_{mn}$
and frequency shifts $\Delta \omega_{mn}$ for an $N$-level system.

\subsection*{Inverting Dephasing Rates}

Using this canonical form, we can determine a set of standard operators
which generate experimentally observed dephasing rates $\Gamma_{mn}$ and
frequencies $\omega_{mn}$(or frequency shifts). The inversion process
relies on the fact that the dephasing rates involving the first $k+1$
levels depend only on the first $k$ dephasing operators, i.e.,
$\Gamma_{12}$ determines the first non-zero element of $V_1$, which
together with $\{\Gamma_{13},\Gamma_{23},\Delta\omega_{23}\}$ then
determines a further three real parameters, and so forth. Hence, it is a
simple matter of iteratively solving a nested set of quadratic
equations, as detailed in the Methods.

Naturally arising in the course of this inversion, a set of constraints
on the allowed dephasing rates and frequency shifts takes the form of
$N-1$ inequalities involving the first $N$ levels
\begin{equation}
  2\Gamma_{1n} - \sum_{\ell=1}^{n-2} |a^{(n)}_\ell|^2 \ge 0, \quad
 \forall n = 2, \ldots, N, 
\end{equation}
where the $a^{(n)}_\ell$ can be expanded in terms of the
$\Gamma_{mn'}$ with $m,n'\le n$.  A more symmetric form of the
constraints is possible, e.g., for $N=3$
\begin{equation*}
2(\Gamma_{12}\Gamma_{23}+\Gamma_{23}\Gamma_{13}+\Gamma_{12}\Gamma_{13})\ge
\Gamma_{12}^2+\Gamma_{23}^2+\Gamma_{13}^2 + \Delta\omega_{23}^2
\end{equation*}
which is reduces to Eq.~(25) in~\cite{SS2004} if $\Delta\omega_{23}=0$.
However, as the number of decoherence rates grows as $N(N-1)/2$ and the
inequalities involve products of $(N-1)$ $\Gamma_{mn}$, there is a
combinatorial explosion in the number of terms in the constraints, which is
why previous attempts to obtain a general form for the constraints 
failed. For example, the four-level constraint contains $22$ terms and the
five-level constraint contains $130$ terms.

\begin{figure}
\includegraphics[width=\columnwidth]{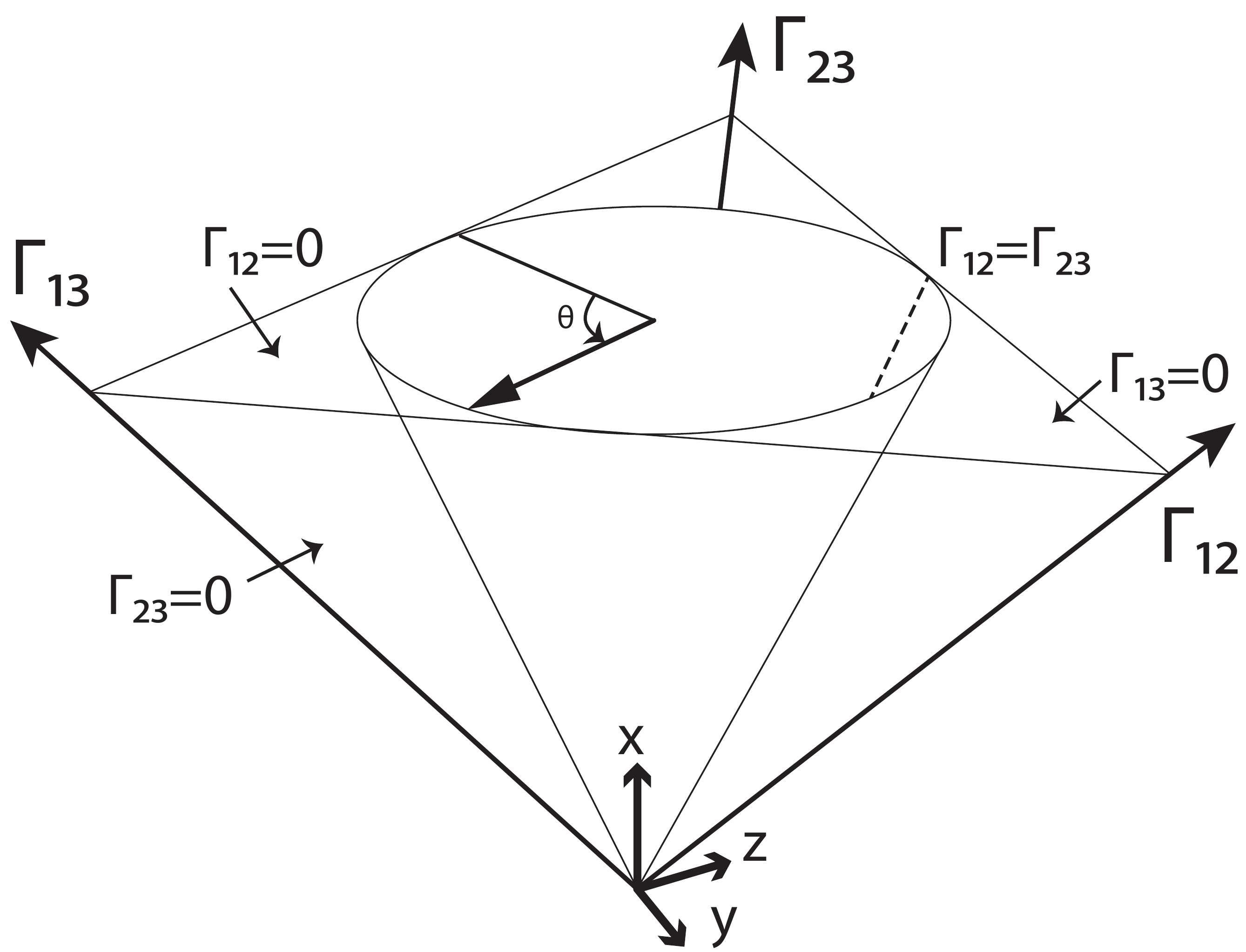}
\caption{Convex cone of allowed dephasing rates for $N=3$ and real
dephasing, i.e., $\Delta\omega_{23}=0$.  The axes are
$x=(\Gamma_{12}+\Gamma_{23}+\Gamma_{13})/\sqrt{3}$,
$y=\tfrac{1}{\sqrt{2}}(\Gamma_{13}-\Gamma_{23})$ and
$z=\sqrt{\tfrac{2}{3}}
\left(\Gamma_{12}-\tfrac{1}{\sqrt{2}}(\Gamma_{13}+\Gamma_{23}) \right)$
and the constraint equation becomes $x^2/2 \ge y^2+z^2$, which defines a
circular cone in the positive octant of in the parameter space of
$\{\Gamma_{12},\Gamma_{23},\Gamma_{13}\}$, tangential to the
$\Gamma_{12}\Gamma_{23}$, $\Gamma_{23}\Gamma_{13}$, and
$\Gamma_{13}\Gamma_{12}$ planes.}  \label{fig:cone}
\end{figure}

These inequalities form a convex cone of allowed dephasing
rates whose boundary is formed by ``hypersurfaces'' defined by
$a^{(k)}_{k-1}=0$ for some $k>1$.  For $N=3$ there is only a single
constraint equation and the convex cone of allowed dephasing rates can
be visualized as shown in Figure~\ref{fig:cone}.

\subsection*{Speed Limits for Entanglement Decay}

The constraints for the decoherence rates and frequency shifts have
important implications for a wide range of physical, chemical and
biological systems where phase relaxation is a dominant process.  A
consequence is the imposition of relative speed limits on the rates at
which coherences can decay, especially in multi-partite systems where
entanglement decay is strictly bounded above by the single qubit
dephasing rates. Dephasing can be spatially correlated, the Markovian
condition only constrains the temporal correlations in the noise. In
general, the dephasing rates can be of a non-local form.

\begin{figure}
\includegraphics[width=\columnwidth]{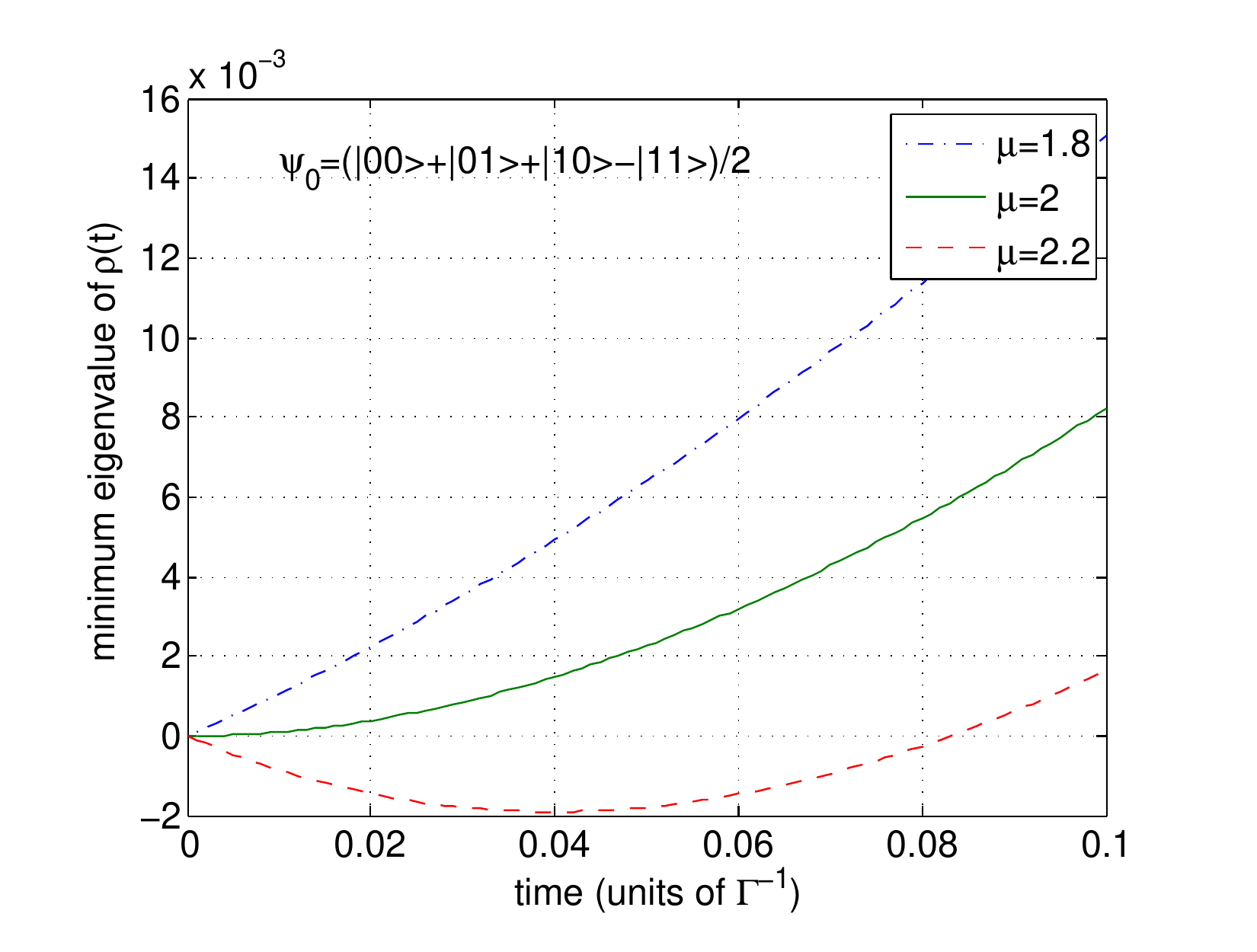}
\caption{Constraint violation leads to non-physical states.  A plot of
the minimum eigenvalue of $\rho(t)$ starting with $\rho_0=
\ket{\Psi_0}\bra{\Psi_0}$ subject to pure dephasing $(H=0)$ with
$\Gamma_{23}=\Gamma_{14}=\mu\Gamma$ for different values of $\mu$ shows
the emergence of negative eigenvalues for $\mu>2$.}  \label{fig2}
\end{figure}

Let us start with two qubits where we label the basis states by
$\ket{1}=\ket{00}$, $\ket{2}=\ket{01}$, $\ket{3}=\ket{10}$ and
$\ket{4}=\ket{11}$.  Assuming that both qubits have the same local
dephasing rate, i.e.
$\Gamma=\Gamma_{12}=\Gamma_{13}=\Gamma_{24}=\Gamma_{34}$, then the
allowed decoherence rates for the non-local coherences $\Gamma_{14}$ and
$\Gamma_{23}$ are determined by $\Gamma$.  The first non-trivial
constraint $(a_{2}^{(3)})^2\ge 0$ gives $0\le \Gamma_{23}\le 4\Gamma$.
The second constraint $(a_{3}^{(4)})^2\ge 0$ leads to
$\Gamma_{23}+\Gamma_{14}\le 4\Gamma$~\footnote{An upper bound of
$4\Gamma$ on the non-local dephasing time was also found in
\cite{YuEberly2002} the Markovian limit for a specific exactly solvable
model of phonon decoherence, in contrast to the Non-Markovian regime,
where much faster entanglement decay was shown to be possible.}. Thus,
to ensure complete positivity of the evolution, the non-local coherences
$\rho_{23}$ and $\rho_{14}$ can decay at most four times as fast as the
local coherences, and the sum of the non-local decay rates can be no
more than $4\Gamma$.  If they are equal
$\Gamma_{23}=\Gamma_{14}=\Gamma_e$ we obtain $\Gamma_e\le 2\Gamma$, and
Fig.~\ref{fig2} demonstrates that violation of the bound leads to
violations of positivity, i.e., non-physical states.


The limits to non-local coherence decay relate to the entanglement
between qubits.  Starting with the maximally entangled Bell state
$\ket{\Psi_0}=\tfrac{1}{\sqrt{2}}(\ket{00}+\ket{11})$, the
state evolving under pure dephasing
\begin{equation*}
  \rho_{\Psi_0}(t) = \frac{1}{2} \begin{pmatrix} 1 & 0 & 0 & e^{-t\Gamma_{14}}\\
                         0 & 0 & 0 & 0 \\ 0 & 0 & 0 & 0 \\ 
                         e^{-t\Gamma_{14}} & 0 & 0 & 1 \end{pmatrix}
\end{equation*}
has concurrence $C(t)=e^{-t\Gamma_{14}}$~\cite{Wootters2001}, thus
$\Gamma_{14}+\Gamma_{23} \le 4\Gamma$ implies that the concurrence
cannot decay faster than four times the local decoherence rate
$\Gamma$.  Here, the decay of non-local coherences is not lower
bounded, i.e. non-local coherences can survive indefinitely even for
finite local decay rates and in this case the entanglement is
preserved if $\Gamma_{14}=0$, i.e.\ there is no sudden death of
entanglement~\cite{YuEberly2004}.

Alternatively, starting with the maximally entangled two-qubit cluster
state
$\ket{\Psi_{CS}}=\frac{1}{2}(\ket{00}+\ket{01}+\ket{10}-\ket{11})$, we
obtain
\begin{equation*}
 \rho_{CS}(t) = \frac{1}{4} \begin{pmatrix}
           1 & e^{-\Gamma t} & e^{-\Gamma t} & -e^{\Gamma_{14}t} \\
           e^{-\Gamma t} & 1 & e^{-\Gamma_{23}t} & -e^{-\Gamma t} \\
           e^{-\Gamma t} & e^{-\Gamma_{23}t} & 1 & -e^{-\Gamma t} \\
           -e^{-\Gamma_{14} t} & -e^{-\Gamma t} & -e^{-\Gamma t} & -1 
           \end{pmatrix}.
\end{equation*}

In this case the entanglement can decay even if both non-local
dephasing rates vanish, $\Gamma_{14}=\Gamma_{23}=0$, in which
case the concurrence satisfies $2C_1(t)=|e^{-\Gamma t}+1|-|e^{-\Gamma
  t}-1|$, which tends to zero as $t\to\infty$.  
If one of the two
non-local concurrences is $4\Gamma$ and the other is $0$, e.g.,
$\Gamma_{14}=4\Gamma$, $\Gamma_{23}=0$, the concurrence similarly
decays asympotically but faster.  When the non-local coherences
decay at the same rate $\Gamma_{14}=\Gamma_{23}=2\Gamma$ we have
$C_2(t)=\max\{0,\tfrac{1}{2}(2e^{-\Gamma t}+e^{-2\Gamma t}-1)\}$, and
the concurrence vanishes when $2e^{-\Gamma t}+e^{-2\Gamma t}-1=0$,
i.e. $t_*= -\Gamma^{-1}\log(\sqrt{2}-1)\approx 0.383\Gamma^{-1}$,
i.e., we observe sudden death of entanglement
rather than asymptotic decay.

\begin{figure}
\includegraphics[width=\columnwidth]{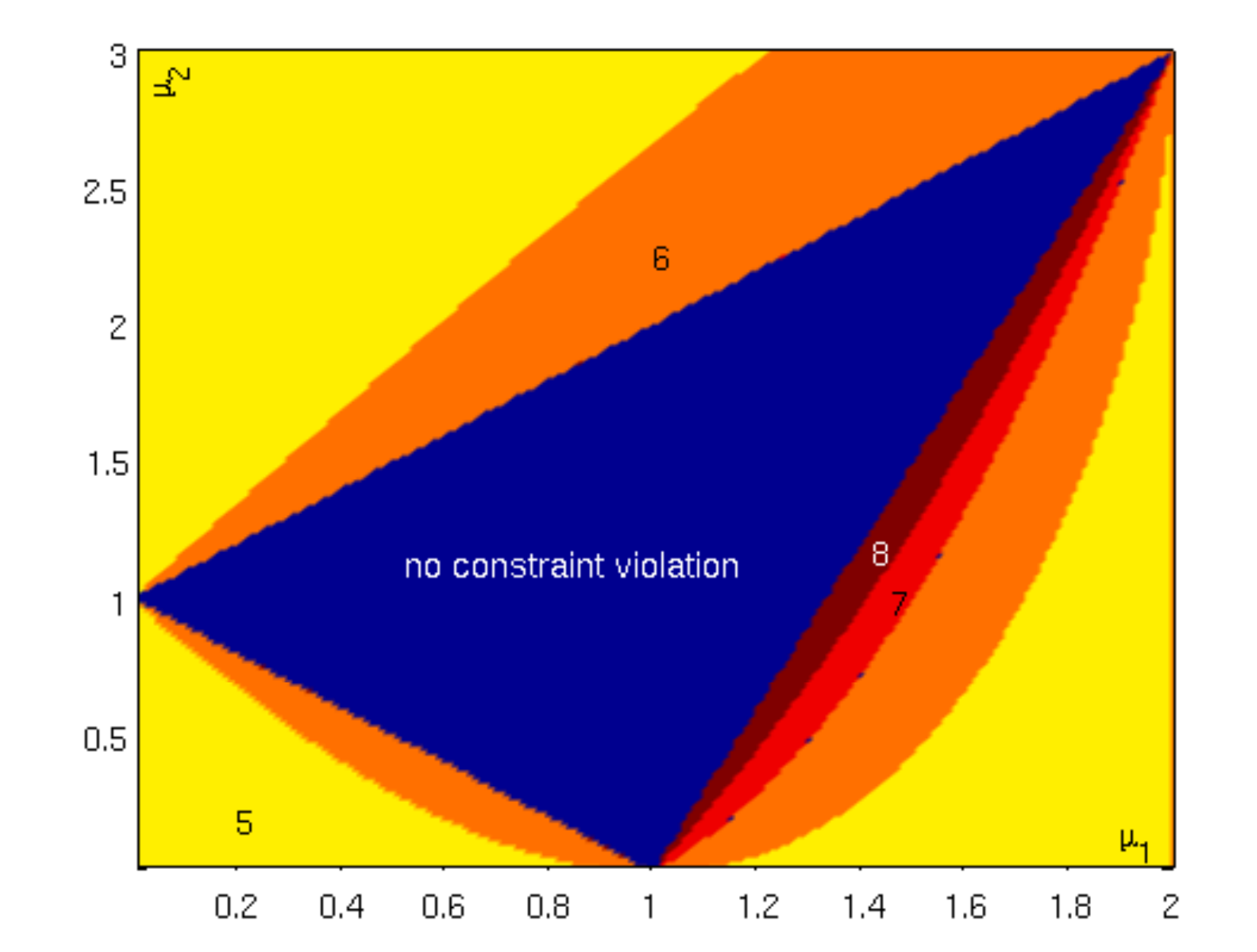}
\caption{Constraint violation map for three qubits with local
  dephasing rates $\Gamma$, two-qubit dephasing rates $\mu_1\Gamma$,
  and three-qubit dephasing rates $\mu_2\Gamma$. For $0\le\mu_{1}\le
  2$ the first four of the eight constraints are satisfied but
  additional constraints may be violated, e.g. in the yellow region,
  constraint 5 is violated. Each subsequent constraint further
  restricts the set of allowed rates.}  \label{fig:const1}
\end{figure}

We can extend this to $n$-qubit systems. Given $n$ qubits with the
same local dephasing rate $\Gamma$, we can simply apply the results
above to any subsystem consisting of two qubits, i.e., the
entanglement between any two qubits in the system cannot decay faster
than $2\Gamma$. For larger systems there are more constraints so in
practice the rate of entanglement decay between any two qubits would
be even more restricted.  For example for a three-qubit system with
have $N=2^3=8$.  Assuming the local dephasing rate for each qubit is
$\Gamma$, and the dephasing rate involving two- and three-qubit
transition terms are $\mu_1\Gamma$ and $\mu_2\Gamma$, respectively,
there are 8 constraints restricting the allowed values for $\mu_1$ and
$\mu_2$.  From the constraints for the two-qubit system we know that
$0\le \mu_1\le 2$, but Fig.~\ref{fig:const1} shows that the set of
$(\mu_1,\mu_2)$ that satisfy all the constraints is much smaller.
Each additional constraint reduces the set of allowed dephasing rates.

\section*{Discussion}

The underlying basis for the dephasing constraints is correlations
between noise acting on different energy levels of the system. The
canonical dephasing operators reflect underlying physical processes
with different correlation properties. For example, a canonical
dephasing operator with a single non-zero element can be interpreted
as the result of the fluctuation of a single energy level. Multiple
non-zero diagonal entries correspond to correlated perturbation of
more than one level. These fluctuations can be correlated across
levels even for Markovian dynamics where the noise statistics are
temporally uncorrelated.

An example of where noise correlation can occur is magnetic field
fluctuations acting on a spin-1 particle where the coupling is of the
form $B_Z S_Z$. This leads to anti-phase perturbations of the $S_Z=\pm
1$ levels and the canonical operator in the basis
$\{\ket{0},\ket{1},\ket{-1}\}$ is $\propto\diag(0,1,-1)$. If the
coupling was instead of the form $B_Z S_Z^2 $, then the canonical
operator would be $\propto\diag(0,1,1)$.

Dephasing not only can lead to exponential damping of the coherences
but also can produce shifts in their frequencies. Not all of these
frequency perturbations can be accommodated by modifying the system
Hamiltonian in general, the residual shifts are intrisic to the
decoherence processes. Whilst pure damping can be generated by phase
diffusion due to random drift of the energy levels (Wiener-Levy
process)~\cite{Wodkiewicz1979}, the frequency shifts can be caused by
phase kicks, or discrete random phase jumps with a poissonian arrival
time, though this also produces additional
damping~\cite{Herzog1995}. Phase kicks occur, for example, by
collisional processes in gases, whereas phase diffusion can be
generated by white noise acting on the energy levels.

It is possible to derive dephasing constraints from physical models of
the noise directly, though deriving the general multi-level
constraints is considerably more difficult than the methods shown
here. However, once the observed dephasing rates have beeen decomposed
into their corresponding canonical set of dephasing operators, we can
assign physical mechanisms by which they occur, and hence perform
system diagnostis or analysis. The ability to identify sources of
dephasing will be vital in producing coherent quantum devices and
improving their performance.

In the context of multi-partite systems, the constraints we have derived
have implications for the preservation of non-local correlations. As the
number of parties increases, the decay of the non-local coherences
becomes constrained even more by the local dephasing rates. This
reflects the general robustness of the non-local correlations in
multi-partite systems~\cite{MultiPartCorrelationRobustness}.
Conversely, there are suggestions that dephasing can play a positive
role in biological processes~\cite{Plenio2008,Plenio2008}. Dephasing has
been mooted to enhance the transport of energy networks such as
photosynthetic harvesting complexes. In such systems, measurement and
analysis of the dephasing may illuminate these processes and lead to
better energy collection devices.

\section*{Methods}
 
\noindent\textbf{Canonical dephasing operators.}  
We start with the Lindblad master equation (LME) for Markovian open
quantum system evolution
\begin{equation}
  \label{eq:LME}
  \dot{\rho}(t) = -i (H\rho(t)-\rho(t)H) + \LL_D(\rho(t)),
\end{equation}
where $\rho(t)$ is the density operator describing the system state
(defined on the system Hilbert space $\H_s$), $H$ is an effective
Hamiltonian $H$ and the superoperator $\LL_D(\rho)$ takes the form
$\LL_D(\rho)=\sum_k \D[V_k]\rho$ with~\cite{Lindblad1976,GKS1976}
\begin{equation}
  \label{eq:D}
  \D[V_k]\rho 
 = V_k \rho V_k^\dag -\tfrac{1}{2}(V_k^\dag V_k\rho + \rho V_k^\dag V_k).
\end{equation}
for operators $V_k$ on $\H_S$.

The operators $(H,V_k)$ define a pure dephasing process with respect to
the basis $\BB=\{\ket{n}\}_{n=1}^N$ if and only if $H$ and all $V_k$ are
simultaneously diagonal with respect to $\BB$, i.e., we have
\begin{subequations}
\label{eqn:diag}
\begin{align}
  H  &= \sum_n\lambda_n  \Pi_n = \diag(\lambda_n), \quad \lambda_n\in\RR \\
 V_k &= \sum_n\gamma_{nk}\Pi_n = \diag(\gamma_{nk}).
\end{align}
\end{subequations}
This is easy to see since by definition of a pure dephasing process the
populations of the basis states remain constant, and thus each basis
state $\ket{n}$ is a steady state of the system.  This is possible only
if the subspace spanned by each basis state $\ket{n}$ is $V_k$-invariant
for all $V_k$~\cite{SW2010}.  This shows that all $V_k$ must be diagonal
in the chosen basis.  Since $\Pi_n$ is diagonal and diagonal operators
commute we have $\D[V_k](\Pi_n)=0$ for all $n$ and all $k$.  As
$\ket{n}$ is a steady state, i.e., $\dot{\Pi}_n(t)=0$, it also follows
that $-iH\Pi_n+i\Pi_n H=0$ for all $n$.  Inserting this into the general
form of the LME~(\ref{eq:LME}) gives the explicit equation
\begin{equation}
  \dot{\rho}_{mn}(t) = -\left(i\omega_{mn}+\Gamma_{mn}\right) \rho_{mn}(t), 
\end{equation} 
for the evolution of the matrix elements $\rho_{mn}=\bra{m}\rho\ket{n}$
of the density operator, or in integral form (\ref{eq:rhomn}) with
frequencies $\omega_{mn} =\lambda_m-\lambda_n+\Delta\omega_{mn}$ and
dephasing induced frequency shifts and decoherence rates given by
(\ref{eqs:omgGamma}).

Any set of diagonal Lindblad operators $\{V_k\}$ generates pure
dephasing dynamics but the set of Lindblad operators $\{V_k\}$
generating a certain dynamical evolution is not unique.  In particular,
we have \emph{unitary invariance}, i.e., given any set of Lindblad
operators $\{V_k\}$, the set of operators $\{W_j\}$ defined by
\begin{equation}
  \label{eq:oprearrange}
  W_j=\sum_k u_{jk} V_k, 
\end{equation}
where $u_{jk}$ are elements of a unitary matrix, generates the same
dynamics as $\sum_k\D[W_k]\rho=\sum_k\D[V_k]\rho$.  
Furthermore \emph{adding multiples of the identity}, $V_k\to V_k+\alpha
\ONE$, to a Lindblad operator $V_k$ only changes the effective
Hamiltonian
\begin{equation}
  \label{eq:rescale}
  \D[V_k+\alpha \ONE]\rho = 
  \tfrac{1}{2}[\alpha V_k^\dag-\alpha^* V_k,\rho(t)] +\D[V_k]\rho,
\end{equation}
and thus the dynamics is unchanged if we replace $V_k$ by
$V_k+\alpha\ONE$ and $H$ by $H+\tfrac{i}{2}(\alpha V_k^\dag-\alpha^*
V_k)$.  

The invariance of the dephasing dynamics under the two ``gauge
transformations'' (\ref{eq:oprearrange}) and (\ref{eq:rescale}) allows
us to transform any set of dephasing operators $\{V_\ell\}$ into an
equivalent set of dephasing operators in canonical form defined in
Eq.~(\ref{eq:canonical}), which yield the same observable dephasing
rates $\{\Gamma_{mn}\}$ and dephasing shifts $\{\Delta\omega_{mn}\}$,
using Algorithm~\ref{algo1}.  The process is constructive and, using
$a_{nk}$ intead of $a_n^{(k)}$, the key steps can be described as
follows:

(1) Using (\ref{eq:rescale}) we ensure that $a_{1k}=0$ for all $V_k$, 
modifying the Hamiltonian by
\begin{equation}
   \Delta H = \frac{i}{2}\sum_k a_{1k} V_k^\dag - a_{1k}^* V_k
\end{equation}
as necessary.

(2) We replace the Lindblad operators $V_1=\diag(0,a_{21},a_{31},\ldots)$ 
and $V_2=\diag(0,a_{22},b_{32},\ldots)$ with $a_{21}=r_{21} e^{i\phi_{21}}$
and $a_{22}=r_{22} e^{i\phi_{22}}$ by $\{W_1,W_2\}$ with
\begin{subequations}
 \label{eq:trans1}
\begin{align}
  W_1 &= u_{11} V_{1} + u_{12} V_{2} 
       = \diag(0,c,\ldots)\\
  W_2 &= u_{21} V_{1} + u_{22} V_{2}
       = \diag(0,0,*,\ldots) 
\end{align}
\end{subequations}
with the unitary coefficient matrix
\begin{equation}
  u = \frac{1}{c}
      \begin{pmatrix} 
       r_{21}e^{i\phi_{22}} & r_{22} e^{i\phi_{21}}\\ 
       r_{22}e^{-i\phi_{21}} & -r_{21}e^{-i\phi_{22}}
       \end{pmatrix}
\end{equation}
and $c=\sqrt{r_{21}^2+r_{22}^2}$, which is dynamically equivalent to
$\{V_1,V_2\}$ due to (\ref{eq:oprearrange}).

This result allows us to reduce an arbitrary number of parameters,
specified by the non-zero elements of a general set $\{V_k\}$ of
dephasing operators to $N(N-1)/2$ parameters in the canonical form.
Note that the number of free parameters matches the number of dephasing
rates $\Gamma_{mn}$ for an $N$-level system.  The procedure will work,
i.e., to produce a set of canonical dephasing operators that reproduce
the observed dephasing rates and shifts, provided that these satisfy the
positivity constraints.  Furthermore, if the observed dephasing rates
and shifts lead to constraint violations these will be detected and
flagged, and this information can be used to further investigate if the
violations can be explained in terms of uncertainty in the observed
data, e.g., due to measurement errors, or if they are indicative of
processes that would invalidate the Markovian dephasing assumption.

We note that if one has the usual Kossakowski form of Markovian
evolution~\cite{GKS1976}, we can ``diagonalise'' the sets of operators
to arrive at a Lindblad form~\cite{Lindblad1976} where the decoherence
operators are orthogonal and traceless. However, this standard form is
not convenient for inversion, nor does it give much physical insight
into the possible processes leading to dephasing. The canonical form
Eq.~\ref{eq:canonical} decomposes the dephasing into operators
representing correlated level perturbations of orders 1 to N-1.

\begin{KCalgorithm}
\KCin{$W$}{Matrix $(N\times K)$, $k$th column equals 
           diagonal elements of Lindblad operator $V_k$}
\KCout{$V$}{Lower triagonal matrix, columns equals 
            diagonal elements of canonical $V_k$}
\KCcode{CanonicDephasing}{Calculate Canonical Dephasing
Operators}{99}{\columnwidth}{
   \KClet{$R$}{Number of rows of $W$}
   \KClet{$C$}{Number of columns of $W$}
   \KClet{$V$}{$W-$\KCid{ones}$(R,1)*W[1,:]$}
   \KClet{$k$}{$1$}[Running column index]
   \KCfor{$r$}{$2,\ldots, N$}{
      \KClet{$i_1$}{Index 1st nonzero entry of $V[r,k:C]$}
      \KClet{$i_1$}{$i_1+k-1$}[shift index]
      \KCwhile{more than one element of $V[r,k:C]$ non-zero}{
         \KClet{$i_2$}{Index 2nd nonzero entry of $V[r,k:C]$}
         \KClet{$i_2$}{$i_2+k-1$}[shift index] 
         \KClet{$r_1$}{$|V[r,i_1]|$},
         \KClet{$r_2$}{$|V[r,i_2]|$}
         \KClet{$\phi_1$}{\KCid{Phase}$(V[r,i_1])$},
         \KClet{$\phi_2$}{\KCid{Phase}$(V[r,i_2])$}
         \KClet{$n_c$}{$\sqrt{r_1^2+r_2^2}$}
         \KClet{$V[:,i_1]$}{$(r_1 e^{+i\phi_2} V[:,i_1]+ 
                               r_2 e^{+i\phi_1} V[:,i_2])/n_c$}
         \KClet{$V[:,i_2]$}{$(r_2 e^{-i\phi_1} V[:,i_1]- 
                               r_1 e^{-i\phi_2} V[:,i_2])/n_c$}
	 }
      \KCif{$V[r,k:C]$ has non-zero entries}{
         \KClet{$k_0$}{Index of 1st non-zero entry}
         \KClet{$k_0$}{$k_0+k-1$}
         \KClet{$V$}{Swap columns $k$ and $k_0$ of $V$}
         \KClet{$k$}{$k+1$}
	 }
      }
   \KClet{$V$}{Remove $0$ columns of $V$, apply phase corrections}
}
\caption{Canonical Dephasing Operators}\label{algo1}
\end{KCalgorithm}

\section{Acknowledgements}

DKLO acknowledges support from the Quantum Information Scotland network
(QUISCO).  SGS acknowledges funding from EPSRC ARF Grant EP/D07192X/1
and Hitachi.

\end{document}